\documentstyle[aps,twocolumn]{revtex}
\begin{document}
\draft

\title{
Density matrix interpretation of solutions of Lie-Nambu equations
}
\author{Marek Czachor{$^{1}$} and Marcin Marciniak{$^2$}}
\address{
${^1}$Katedra Fizyki Teoretycznej i Metod Matematycznych,
 Politechnika Gda\'{n}ska\\
ul. Narutowicza 11/12, 80-952 Gda\'{n}sk, Poland\\
E-mail: mczachor@sunrise.pg.gda.pl\\
$^2$Instytut Matematyki,
Uniwersytet Gda\'{n}ski,
ul. Wita Stwosza 57, 80-952 Gda\'{n}sk, Poland
}
\maketitle
\begin{abstract}
The spectrum of a density matrix $\rho(t)$ is conserved by a
Lie-Nambu dynamics if $\rho(t)$ is a self-adjoint and
Hilbert-Schmidt solution of a nonlinear triple-bracket equation. 
This generalizes to arbitrary separable (positive- and
indefinite-metric)  Hilbert spaces
the previous result which was valid for finite-dimensional
Hilbert spaces. 
\end{abstract}


\section{State vectors vs. density matrices in nonlinear quantum mechanics}

There exist prejudices concerning nonlinear generalizations
of quantum mechanics. One of them is a belief that any 
generalization must lead to unphysical effects such as a
faster-than-light transfer of information. Although the proofs
of these unphysical phenomena are explicit and {\it mathematically\/}
correct \cite{G1,G2,MC01,MC02} they are based on some {\it physical\/}
assumptions which are unjustified. One of these physically wrong
elements is a naive use of the projection postulate. Today we
understand that if the dynamics is nonlinear we are not allowed
to simply project a solution on some direction in a Hilbert
space. One of the reasons is that a projection of a solution is
in general not a solution of a nonlinear Schr\"odinger equation.
A more subtle argument is provided by the notion of nonlinear
gauge transformations introduced by Doebner and Goldin \cite{DG,GAG}
and developed by the Clausthal school \cite{C}. It is clear that
there exists a class of nonlinear Schr\"odinger equations which
are obtained by a nonlinear gauge transformation from an
ordinary linear Schr\"odinger equation. They not only give the
same probability density in position space but also may look
``truely" nonlinear (there exist nonlinear gauge transformations
that simply add a nonlinear term but do not alter the kinetic
and potential parts in a Hamiltonian). Obviously if the
nonexistence theorems were true such ``nonlinear" equations
would have to lead to unphysical effects. But the point is that
they do not lead to {\it any\/} new effects since, by definition,
they are physically fully equivalent to to the linear theory.
Therefore the nonexistence theorems must contain some elements
which are physically wrong. From the perspective of the
nonlinear gauge transformations it is clear that one of them is a wrong
use of the projection postulate. Let us note,
however, that the
explicit example discussed in \cite{MC01} is not based on this postulate
(and thus is not equivalent to the examples given in
\cite{G1,G2}; a simple argument shows also that the ``telegraph"
discussed in \cite{MC01} works in the opposite direction than
those from \cite{G1,G2}). An element which is physically
wrong here is the wrong way of describing composite systems.
This was clarified by Polchinski \cite{P} and Jordan \cite{J}.
The latter work was based on a density matrix reformulation of
Weinberg's nonlinear quantum mechanics \cite{W,Bona}. 

Density matrices play in nonlinear quantum mechanics a role
which is somewhat ambiguous. On the one hand, one of the earliest
attempts of formulating a general nonlinear framework for
quantum mechanics was Mielnik's ``convex formalism" \cite{M}.
Its main idea was to keep a figure of states convex and derive a
probability interpretation in terms of its global geometric
properties. From this perspective the density matrices might be even
more fundamental than state vectors. On the other hand, however,
all works that start from pure states and nonlinear
Schr\"odinger equations lead to difficulties when it comes to
``mixtures". The difficulties are so deep that some authors tend
to reject the very notion of a density matrix in a nonlinear
context \cite{private}, although different proposals of
combining mixtures with nonlinearity of pure states exist in
the literature (cf. \cite{KJ,B}).
 
A nonlinear extension of
quantum mechanics based on a triple bracket Nambu-type
generalization of the Liouville-von Neumann equation (cf.
\cite{BM}) proposed by one of us \cite{MC1,MC2,MC3} starts from
a completely different perspective. 
The idea is to find a general scheme which on one hand 
includes the linear and Weinberg-Bona-Jordan cases and on the other
leaves some room for nonlinear generalizations that do not
use nonlinear Hamiltonians. Such a starting point is
motivated by ambiguities 
in probability interpretation caused by the notion of eigenvalue
of a nonlinear operator \cite{MCpra}. The Lie-Nambu scheme
proved very powerful and elegant and has, in our oppinion, several
advantages over the standard paradigm of nonlinear Schr\"odinger
equations. The density matrices play in this formalism a
fundamental role. 
Still the basic question of an interpretation of solutions
$\rho(t)$ as density matrices was not fully clarified in the
earlier work. The Theorem~5 discussed in \cite{MC2} worked
essentially in finite dimensional cases whereas the generic
infinite dimensional Hilbert space problem was left open.
In this paper we give an alternative proof which generalizes
this theorem to infinite dimensional Hilbert spaces. 
We consider also the indefinite-metric case which is of some
interest for a relativistic theory.

\section{Triple-bracket equations and spectral properties of
their solutions}

Consider a one-parameter family $\rho=\rho(t)$, $t\in \bbox R$, 
of Hilbert-Schmidt
self-adjoint operators acting in a separable Hilbert space
and satisfying the Lie-Nambu equation 
\begin{eqnarray}
i \dot \rho_a=\{\rho_a,H,S\}.\label{1}
\end{eqnarray}
Here $\rho_a:=\rho_{AA'}(\bbox a,\bbox a')$ are components of
$\rho$ in some basis, $A$ and $A'$ are discrete (say, spinor)
indices and $\bbox a$, $\bbox a'$ the continuous ones
(coresponding to, say, position or momentum). 
The dot represents a derivative with respect to the parameter
$t$. In nonrelativistic case this is just an ordinary time. In the
relativistic case the meaning of $t$ depends on a formalism ($t$
is time in some reference frame in \cite{MC2}, and a ``proper
time" in the off-shell formulation given in \cite{MC3}). 
$H=H(\rho)$ is any (functionally) differentiable functional of
$\rho$ and
$S=S(\rho)=S\big(C_1(\rho),\dots,C_k(\rho),\dots\big)$ is
differentiable in $C_n(\rho)$ (see Appendix~\ref{A2}). 
We assume the
following summation convention for the composite indices $a$: A
contraction of two composite indices means simultaneous 
summation over the
discrete indices and integration (with respect to an appropriate
measure) of the continuous ones. The triple bracket
itself is defined as 
\begin{eqnarray}
\{F,G,H\}=
\Omega_{abc}\frac{\delta F}{\delta\rho_a}
\frac{\delta G}{\delta\rho_b}
\frac{\delta H}{\delta\rho_c}
\end{eqnarray}
where $\Omega_{abc}$ are structure constants of an
infinite-dimensional Lie algebra which also depends on the
model \cite{MC2,MC3} (see Appendix~\ref{A1}). The indices in
$\Omega_{abc}$ can be raised and lowered by a metric discussed
in detail in \cite{MC2,MC3}. The metric is well defined for both
finite- and infinite-dimensional Lie algebras and is not
equivalent to the Cartan-Killing one (the latter does not exist in the
infinite-dimensional case).

Before we proceed with the main theorem we shall first prove a
few useful technical results. 

{\it Lemma 1\/}:
Let $\{ p_k\}_{k=1}^\infty$ be a sequence of nonnegative numbers
such that $p_1\geq p_2\geq p_3\geq\ldots$ and the series
$\sum_{k=1}^\infty p_k$ is convergent. Then
$$\lim_{m\to\infty}\left(\sum_{k=1}^{\infty}p_k^m\right)^{\frac{1}{m}}=p_1
.$$ 

\noindent
{\it Proof.} Using the three sequences theorem one immediately generalizes
the standard proof known for finite sequences. $\Box$

{\it Lemma 2\/}:
Let $\{ p_k\}_{k=1}^\infty$, $\{ q_k\}_{k=1}^\infty$ be two sequences
fulfiling the assumptions of the above lemma.
Suppose that for every $m\in {\bf N}$
$$\sum_{k=1}^\infty p_k^m =\sum_{k=1}^\infty q_k^m .$$
Then $p_k=q_k$ for every $k=1,2,\ldots$.

\noindent
{\it Proof.}
The equality $p_1=q_1$ follows directly from Lemma 1.
The proof is completed by induction.$\Box$

{\it Lemma 3\/}:
Let $\{ q_k\}_{k=1}^\infty$, $q_k\in{\bf R}$, be an arbitrary
sequence and 
$\{ p_k\}_{k=1}^\infty$ satisfy the assumptions of
Lemma~1.
Suppose that for every $m\in {\bf N}$
$$\sum_{k=1}^\infty p_k^m =\sum_{k=1}^\infty q_k^m .$$
Then $\{ p_k\}_{k=1}^\infty=\{ q_k\}_{k=1}^\infty$ up to permutation.

\noindent
{\it Proof.} 
Define two sequences
$\{ \tilde p_k\}_{k=1}^\infty$,  $\{ \tilde q_k\}_{k=1}^\infty$, 
where $\tilde p_k=p_k^2$, $\tilde q_k=q_k^2$. 
$\{ \tilde q_k\}_{k=1}^\infty$ is absolutely convergent so we
can rearrange its terms in such a way that the rearranged
sequence satisfies assumptions of Lemma~1. Assume this done. 
Lemma~2 implies that $|q_k|=p_k$, for any $k$. 
Let $Q_+$
be the subset of positive elements of $\{
q_k\}_{k=1}^\infty$. $Q_+$ is non-empty since otherwise
all elements of $\{ q_k\}_{k=1}^\infty$ would be $\leq 0$ which
contradicts the assumption that 
$\sum_{k=1}^\infty q_k=\sum_{k=1}^\infty p_k >0.$ (We exclude
the trivial case where all $p_k=0$.)
Therefore for any $q_k\in Q_+$ there exists $p_k$ such that $p_k=q_k$.
Subtracting these elements from both sides of 
$$\sum_{k=1}^\infty p_k^m =\sum_{k=1}^\infty q_k^m $$
we get for $m=1$
$$\sum_{{\rm remaining}\,p_k} p_k =\sum_{q_k\leq 0} q_k $$
which can hold if and only if all such $q_k=0$. Therefore all
$q_k\geq 0$ and $p_k=q_k$ for any $k$.$\Box$

\medskip
We are interested in solutions of (\ref{1}) where $\rho(t)$ are
Hilbert-Schmidt self-adjoint operators acting in a separable
Hilbert space. The following theorem states
that spectrum of $\rho(t)$ is conserved by the Lie-Nambu dynamics.

{\it Theorem 4\/}:
Let $\rho(t)$ be a Hilbert-Schmidt self-adjoint 
solution of (\ref{1}) whose spectrum is ${\rm sp\,}\rho(t)=
\{\lambda_k(t)\}_{k=1}^\infty$.
If $\{\lambda_k(0)\}_{k=1}^\infty$ satisfies assumptions of Lemma~1 then 
$\lambda_k(t)=\lambda_k(0)$ for any $t$.

\noindent
{\it Proof\/}: 
For any $t$ the solution can be written as 
$\rho_a(t)=\sum_{k=1}^{\infty}\lambda_k(t)\phi^k_A(t,\bbox a)
\bar \phi^k_{A'}(t,\bbox a')$. 
According to Theorem~4 in \cite{MC2} the
functional $C_n(\rho)$, $n\in {\bf N}$, (see Appendix \ref{A2}) is time
independent. Therefore 
\begin{eqnarray}
\sum_{k=1}^{\infty}(\eta^{kk})^n\lambda_k(t)^n
=
\sum_{k=1}^{\infty}(\eta^{kk})^n\lambda_k(0)^n\label{C_n}
\end{eqnarray}
for any $t$ and $n$ ($\eta^{kk}$ includes the indefinite
metric case, see Appendix~\ref{A2}). 
Lemma~3 implies that the sequences 
$\{\lambda_k(t)\}_{k=1}^\infty$ and 
$\{\lambda_k(0)\}_{k=1}^\infty$ are identical up to permutation
if the metric $\eta^{kl}$ is
positive definite. In the indefinite metric case we define 
$\tilde \lambda_k(t)=\lambda_k(t)^2$.
(\ref{C_n}) implies 
\begin{eqnarray}
\sum_{k=1}^{\infty}\tilde \lambda_k(t)^n
=
\sum_{k=1}^{\infty}\tilde \lambda_k(0)^n\nonumber.
\end{eqnarray}
Lemma~3 again implies that the sequences 
$\{|\lambda_k(t)|\}_{k=1}^\infty$ and 
$\{\lambda_k(0)\}_{k=1}^\infty$ are identical up to permutation.
Finally continuity in $t$ means that $\lambda_k(t)=\lambda_k(0)$
in both cases.$\Box$

\section{Convexity principle: An example}

A nonlinearly evolving density matrix cannot satisfy an ordinary
convexity principle. Still, being a Hilbert-Schmidt operator it
can be spectrally decomposed and the spectral projectors can be
regarded as its ``pure state components".  This is justified by
the fact that the spectrum of $\rho(t)$ is time-independent. To
see how this works consider a simple example. The example
simultaneously illustrates the peculiarity of the triple-bracket
formalism: An existence of nonlinearities that become invisible
on pure states. 
Let $H(\rho)={\rm Tr\,}(h\rho)$ where $h$ is 
a $2\times 2$ Hermitian matrix, $\rho
=\rho_0 \bbox 1 +\bbox \rho \cdot\bbox \sigma$, and
\begin{eqnarray}
S(\rho)=\frac{2}{3}
\big({\rm Tr\,}(\rho){\rm Tr\,}(\rho^3)\big)^{1/2}.\label{S_3}
\end{eqnarray}
(\ref{S_3}) is the ``entropy" $S_3$ given by Eq.~(75) in
\cite{MC2}. The Lie-Nambu equation (\ref{1}) is now equivalent
to the matrix equation 
\begin{eqnarray}
i\dot\rho=\Big[\frac{{\rm Tr\,}(\rho)}{{\rm Tr\,}(\rho^3)}\Big]^{1/2}[h,\rho^2].
\end{eqnarray}
Its solution normalized by ${\rm Tr\,} \rho=1$ is
\begin{eqnarray}
\rho(t)=\frac{1}{2}\bbox 1+
\exp\Bigg[\frac{-iht}{\sqrt{\frac{1}{4}+3\bbox
\rho^2}}\Bigg] \bbox \rho\cdot\bbox \sigma
\exp\Bigg[\frac{iht}{\sqrt{\frac{1}{4}+3\bbox
\rho^2}}\Bigg].\label{sol}
\end{eqnarray}
For $h=E\sigma_1$ and $\bbox \rho\cdot\bbox
\sigma=\frac{\varepsilon}{2}\sigma_3$ 
we find $\lambda_1=(1+\varepsilon)/2$, $\lambda_2=(1-\varepsilon)/2$, and
\begin{eqnarray}
\phi_A^1=
\left(
\begin{array}{c}
\cos \omega(E,\varepsilon)t\\
-i\sin \omega(E,\varepsilon)t
\end{array}
\right),
\quad
\phi_A^2=
\left(
\begin{array}{c}
-i\sin \omega(E,\varepsilon)t\\
\cos \omega(E,\varepsilon)t
\end{array}
\right),\nonumber
\end{eqnarray}
where $\omega(E,\varepsilon)=2E/\sqrt{1+3\varepsilon^2}$. Notice that the
vectors $\phi_A^k$, $k=1,2$, depend on $\lambda_k$. In the linear
case we can solve equations for orthogonal pure states and then
form their convex combinations with coefficients $\lambda_k$,
which in no way affects the form of the pure states that form
the mixture. In the nonlinear case the ``pure state" components
of the mixture {\it do\/} depend on the coefficients
$\lambda_k$ \cite{MC3}. The dynamics of
$\rho(t)$ is nonlinear even though the Hamiltonian is given by the linear
operator $h$. A possibility of introducing nonlinearities
without modifications of an algebra of observables is one of the
important differences between the Lie-Nambu formalism and
nonlinear Schr\"odinger equations. For $\bbox \rho^2=1/4$ the
density matrix is a projector and its dynamics is linear. 

Now consider a solution of
(\ref{1}) which at $t=0$ is a convex combination 
\begin{eqnarray}
\rho(0)=p_1 \rho_1(0) + p_2 \rho_2(0)
\end{eqnarray}
of two not necessarily mutually orthogonal density matrices.
Let $\rho_1(0)=\rho_{10} \bbox 1 +\bbox \rho_1 \cdot\bbox
\sigma$  and 
$\rho_2(0)=\rho_{20} \bbox 1 +\bbox \rho_2 \cdot\bbox
\sigma$. The solution we look for is  given by (\ref{sol})
but with $\bbox \rho=p_1 \bbox \rho_1 + p_2 \bbox \rho_2$. 
The Hilbert-Schmidt vectors $\phi_A^k$ depend now not only on the
eigenvalues of $\rho_1(0)$, $\rho_2(0)$, but also on $p_1$ and
$p_2$. This implies also that the dynamics of the density matrix
can be written here as
\begin{eqnarray}
\rho(t)=p_1 U\rho_1(0)U^{\dag} + p_2 U\rho_2(0)U^{\dag}
\end{eqnarray}
where $U=U(t,\rho_1(0),\rho_2(0),p_1,p_2)$ is unitary but
parametrized by the initial condition.

Typically a nonlinear evolution is a result of a mean-field-type
averaging procedure. The example shows that there may exist
another mechanism: Nonlinearity via an entanglement \cite{M2}. Indeed a
subsystem may start to evolve nonlinearly if its reduced density
matrix evolves from a pure state to a mixture, and this happens
whenever the subsystem gets entangled with another one.

\section{Positivity vs. complete positivity}

We have shown that self-adjoint Hilbert-Schmidt 
solutions of nonlinear Lie-Nambu equations are positive for
$t\neq 0$ if they are positive at $t=0$. 
Typically it is assumed that density matrices should be
described by {\it completely\/} positive maps. A physical
motivation behind complete positivity is the problem of
extension of dynamics from subsystems to composite systems
\cite{Davies}: One requires that a dynamics of a system
described by $\rho$ should allow to treat this system as a
trivially embedded subsystem of a bigger one. The evolution of
the bigger system should preserve positivity of {\it its\/}
density matrix. If this is the case, and if for any $t$ the 
density matrix of the composite system is positive independently
of the dimension of the system we have added, 
then the dynamics of the original system is said to be
completely positive. One additional technical assumption one
makes is finite dimensionality of the system one adds. 

In the standard analysis of completely positive maps one assumes the
maps are linear. But the dynamics we have discussed in this
Letter is nonlinear. It can be shown that the definition of a
nonlinear completely positive map introduced in mathematical
literature \cite{AC,A} and applied to Hartree-type equations in
\cite{WAM,AM} is physically incorrect. This problem is analyzed
elsewhere \cite{MCMKprl,MCprl,MCijtp}. From the point of view of this Letter
it is sufficient to note that there exists a large class of
nonliner triple-bracket equations that do not lead to any
problem with extension of dynamics from subsystems to composite
systems. They satisfy all {\it physical\/} requirements
typically associated with the notion of a
completely positive map. Quite surprisingly, they are
not in the form one takes as a departure point for the
discussion of complete positivity of nonlinear maps in the
mathematical literature.

\bigskip
We gratefully acknowledge discussions we had on the subject
with M.~Kuna, P.~Horodecki, G.~A.~Goldin and K.~Jones. The work
of M.~C. is a part of the Polish-Flemish project 007. 

\section{Appendices}

\subsection{Structure constants}
\label{A1}

Consider a Hilbert space of vectors $\psi_\alpha$ where the
abstract index $\alpha$ can be discrete, continuous, or
composite \cite{MC2}. Denote the scalar product by
$\langle\phi,\psi\rangle =\omega^{\alpha\alpha'}\phi_\alpha
\bar\psi_{\alpha'}$. The tensor $\omega^{\alpha\alpha'}$ is in
general a distribution whose inverse is $I_{\alpha\alpha'}$. By
the inverse it is meant that 
\begin{eqnarray}
\omega^{\alpha\alpha'}I_{\alpha\beta'}&=&\delta_{\beta'}{^{\alpha'}}\\
\omega^{\alpha\alpha'}I_{\beta\alpha'}&=&\delta_{\beta}{^{\alpha}}
\end{eqnarray}
where the $\delta$'s mean the Kronecker or Dirac deltas, or
their products. In a Hamiltonian formulation of
Schr\"odinger-type equations $\omega^{\alpha\alpha'}$ and 
$I_{\alpha\alpha'}$ play the roles of a symplectic form and a
Poisson tensor, respectively \cite{MC2,MC3}. 
The structure constants are
\begin{eqnarray}
\Omega{^a}_{bc}&=&\delta_{\beta'}{^{\alpha'}}\delta_{\gamma}{^{\alpha}}
I_{\beta\gamma'}-
\delta_{\gamma'}{^{\alpha'}}\delta_{\beta}{^{\alpha}}
I_{\gamma\beta'}\\
\Omega_{abc}&=&I_{\alpha\beta'}I_{\beta\gamma'}
I_{\gamma\alpha'}-
I_{\alpha\gamma'}I_{\beta\alpha'}
I_{\gamma\beta'}\\
\Omega^{abc}&=&-\omega^{\alpha\beta'}\omega^{\beta\gamma'}
\omega^{\gamma\alpha'}+
\omega^{\alpha\gamma'}\omega^{\beta\alpha'}
\omega^{\gamma\beta'}.
\end{eqnarray}
Notice the spinor-type convention we use.
Different equations (nonrelativistic Schr\"odinger, positive-metric
Dirac, off-shell Dirac, Bargmann-Wigner) correspond to different
Hilbert spaces, $\omega$'s and $I$'s but the form of the
structure constants is always the same. The indices are raised
and lowered by metric tensors defined below.

\subsection{Metric tensors and Casimir invariants}
\label{A2}
We define higher-order metric tensors by
\begin{eqnarray}
g^{a_1\dots a_n}
&=&\omega^{\alpha_1\alpha'_n}\omega^{\alpha_2\alpha'_1} 
\omega^{\alpha_3\alpha'_2}\dots
\omega^{\alpha_{n-1}\alpha'_{n-2}}
\omega^{\alpha_n\alpha'_{n-1}}\label{g}\\ 
G_{a_1\dots a_n}
&=&I_{\alpha_1\alpha'_n}I_{\alpha_2\alpha'_1} 
I_{\alpha_3\alpha'_2}\dots
I_{\alpha_{n-1}\alpha'_{n-2}}
I_{\alpha_n\alpha'_{n-1}}\label{G}\\
&=&
g_{a_n\dots a_1}.
\end{eqnarray}
For $n=1$ we get just the symplectic form and the Poisson
tensor. For $n=2$ we obtain the metric tensor on the Lie
algebra --- it is this tensor that lowers and raises the indices in
the structure constants and (\ref{g}), (\ref{G}). 
For higher $n$'s the tensors define
higher order Casimir invariants of the Lie-Nambu bracket
\begin{eqnarray}
C_n(\rho)=g^{a_1\dots a_n}\rho_{a_1}\dots\rho_{a_n}=
g_{a_1\dots a_n}\rho^{a_1}\dots\rho^{a_n}
\end{eqnarray}
where $\rho_a=\rho_{\alpha\alpha'}$. Let 
$
\rho_a=
\sum_{k=1}^\infty \lambda_k\phi^k_\alpha\bar \phi^k_{\alpha'}
$
be the Hilbert-Schmidt decomposition of $\rho$. The vectors
$\phi^k_\alpha$ are orthonormal  
($\omega^{\alpha\alpha'}\phi^k_\alpha\bar
\phi^l_{\alpha'}=\eta^{kl}=\eta^{kk}\delta^{kl}$; $\eta^{kk}=\pm 1$ if 
the metric is not positive definite) which implies that 
$
C_n(\rho)=\sum_{k=1}^\infty (\eta^{kk})^n\lambda_k^n={\rm Tr\,} (\rho^n).
$
The metric tensors are therefore an abstract index counterpart of trace.

\end{document}